# Fabrication of quantum Hall *p-n* junction checkerboards

Dinesh K. Patel[1,2], Martina Marzano[1,3], Chieh-I Liu[1,4], Mattias Kruskopf[1,5,6], Randolph E. Elmquist[1], Chi-Te Liang[2], and Albert F. Rigosi[1]

[1]Physical Measurement Laboratory, National Institute of Standards and Technology (NIST), Gaithersburg, MD 20899, United States of America
[2]Department of Physics, National Taiwan University, Taipei 10617, Taiwan
[3]Istituto Nazionale di Ricerca Metrologica, Torino 10135, Italy
[4]Department of Chemistry and Biochemistry, University of Maryland, College Park, MD 20742, United States of America
[5]Joint Quantum Institute, University of Maryland, College Park, MD 20742, United States of America
[6]Electricity Division, Physikalisch-Technische Bundesanstalt, Braunschweig 38116, Germany

E-mail: afr1(at)nist(dot)gov

**Abstract**

Measurements of fractional multiples of the $\nu = 2$ plateau quantized Hall resistance ($R_H \approx$ 12906 Ω) were enabled by the utilization of multiple current terminals on millimetre-scale graphene *p-n* junction devices fabricated with interfaces along both lateral directions. These quantum Hall resistance checkerboard devices have been demonstrated to match quantized resistance outputs numerically calculated with the LTspice circuit simulator. From the devices' functionality, more complex embodiments of the quantum Hall resistance checkerboard were simulated to highlight the parameter space within which these devices could operate. Moreover, these measurements suggest that the scalability of *p-n* junction fabrication on millimetre or centimetre scales is feasible with regards to graphene device manufacturing by using the far more efficient process of standard ultraviolet lithography.

## 1. Introduction

Graphene has been demonstrated to be a versatile material since its discovery [1-4], with applications spanning numerous interdisciplinary subfields of study. More specifically, devices fabricated from graphene exhibit desirable transport properties while operating in the quantum Hall regime [5-14]. Coupled with the physics behind *p-n* junctions (*pn*Js), corresponding devices can enable advances in fields like quantum Hall metrology [15-16]. Moreover, the demonstration of functional *pn*Js on millimetre scales and above may offer promising solutions for problems in electron optics, charge density waves, superconductivity, and photodetection [17-24]. Though various linear Hall bar *pn*J devices have been fabricated, most devices have sizes of the order 100 μm or smaller due to the typical requirement of top-gating, whose effectiveness drastically reduces with size owing to the higher probability of current leakage. This size limitation has recently been lifted, as has the requirement of using a top gate to modulate the carrier density in graphene [25-28]. Non-conventional geometries of *pn*J devices have not yet been explored since fabrication methods were still quite intricate and may have posed challenges in the past.

This work reports details on the fabrication and testing of millimetre-scale epitaxial graphene (EG) *pn*J devices with specific junction geometries that resemble simple checkerboards. Subsequent measurements of those devices in the quantum Hall regime were performed to compare experimental quantized resistances to those obtained with the LTspice circuit simulator [29-30]. Since all epitaxial graphene regions exhibit a resistance of $R_K = \frac{h}{2e^2}$ at sufficiently high magnetic flux density (where the von Klitzing constant $R_K$ is defined as containing the Planck constant $h$ and the elementary charge $e$), all non-conventional quantized resistances presented herein will have resulted from the use of several contact terminals as sources or drains [25-28]. The advantage of these complex configurations is that a whole set of quantized resistances becomes accessible and can provide overwhelming evidence of device functionality in addition to being a future application for quantum-based electrical standards.

Specific applications of general checkerboard devices, much like those that will be demonstrated and proposed, also include the construction of a multi-interfaced, two-dimensional Dirac fermion microscope [31], custom programmable quantized resistors [32], and mesoscopic valley filters [33]. It is therefore also important to verify these checkerboard devices as functional so that their fabrication methods can be justified for use in other applications.

## 2. Experimental and Numerical Methods

### 2.1 Graphene growth and device fabrication

EG was grown on a 2.7 cm × 2.7 cm square of single-crystal SiC that was diced from a 4H-SiC(0001) wafer (CREE) [see Notes]. The various procedures for preparing any pieces for growth are detailed in other works [34-37]. The growths were performed in an argon environment at 1900 °C using a graphite-lined, resistive-element furnace from Materials Research Furnaces Inc. [see Notes] with heating and cooling rates of approximately 1.5 °C/s.

After inspecting the growths using confocal laser scanning (CLSM) and optical microscopy to verify monolayer homogeneity, Pd and Au layers were deposited prior to performing any fabrication processes. These layers protect the EG from any organic contamination [34-37]. Some devices had NbTiN deposited as electrical contacts for improved performance [38]. After the devices were completed, each one underwent functionalization treatment with Cr(CO)$_6$. At temperatures as low as 130 °C, the carbonyl compound sublimates to become Cr(CO)$_3$ and attaches to the surface of EG by organometallic covalent chemisorption [39-41]. This treatment was performed in a home-built vacuum chamber and offers carrier density uniformity with default values on the order of $10^{10}$ cm$^{-2}$. Basic annealing was used to achieve a high level of control over the desired carrier density [42].

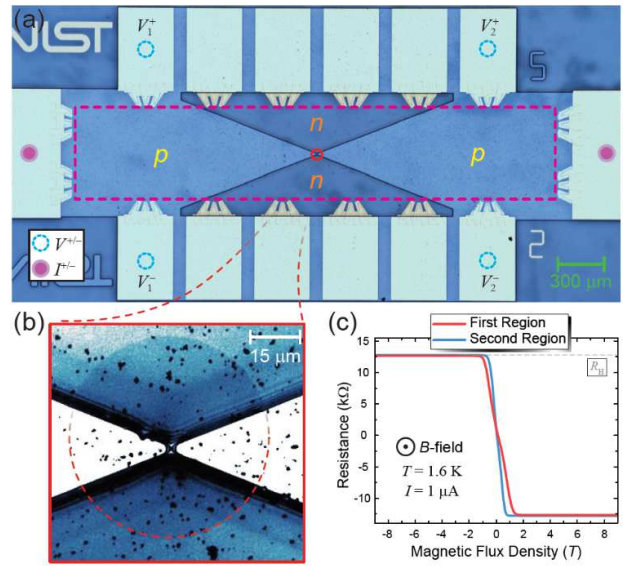

Fig. 1. (a) Optical image of a checkerboard device after fabrication. The pink dashed line outlines the perimeter of the EG. Pink dots and blue circles indicate an example set of current and voltage terminals, respectively (see (c)). A red circle in the center of the device marks the region of EG removed to give each of the four regions a well-defined boundary. (b) A magnification of the red circle in (a) reveals a CLSM image of the intersection of all four regions. The contrast has been optimized such that the EG is noticeably absent in the center circle. (c) Two p-type regions were characterized after adjustment to verify the conversion of the typical n-type doping inherent to EG.

### 2.2 Adjustment of the carrier density

When one fabricates millimetre-scale pnJ devices, it is crucial to ensure that large adjacent regions can accommodate a unipolar value of the carrier density, are opposite in charge carrier sign, and are separated by a sufficiently narrow pnJ to enable Landauer-Büttiker edge state propagation and equilibration [5-9, 43-47].

To that end, an additional set of fabrication processes was implemented using S1813 photoresist intended for protecting n-type regions (inherent due to the buffer layer [48]) and using PMMA/MMA and ZEP520A photoresists as capping materials for regions intended to be p-type [25-26, 49]. The PMMA/MMA acts as a mediation layer for ZEP520A, a common electron beam resist with photoactive properties. The entire device was then exposed to ultraviolet light (254 nm) for anywhere between 10 h – 15 h to allow electrons to be transferred from the EG to the ZEP520A. Regions protected by S1813 did not experience significant changes in the carrier density but still required an annealing process of approximately 25 min (with $T$ = 350 K) to obtain an electron density of about $10^{11}$ cm$^{-2}$. An example optical image of a completed device is shown in Fig. 1 (a). A magnification of the center of the device in Fig. 1 (b) shows the importance of removing a small amount of EG to ensure that all four regions are topologically distinct. Without its removal, the risks remain high that during fabrication, two separate regions would merge, changing the inherent geometry of the device.

Though all *n*-type regions were adjusted by the aforementioned annealing process (owing to its functionalization treatment) [42], verifying the *p*-type regions of the devices required additional measurements since the ultraviolet light exposure is not as well-quantified in the literature. For the example device in Fig. 1 (a), a simple Hall measurement was performed after both the exposure and annealing, with the pink dots as the current terminals and the blue circles as both sets of voltage terminals. The two Hall measurements are shown in Fig. 1 (c), where the carrier density was observed to be both *p*-type and of sufficient quality such that the quantized plateau at $\nu = 2$ was measurable. Measurements were performed at $T = 1.6$ K and $-9$ T $\leq B \leq 9$ T.

*2.3 LTspice simulations*

The electronic circuit simulator LTspice was used for numerically predicting the behavior of both the experimental EG checkerboard devices as well as the proposed ones [29-30]. The circuit was composed of interconnected quantized regions that were modeled as ideal clockwise (CW) or counterclockwise (CCW) *k*-terminal quantum Hall effect elements, depending on whether one modeled *p*-type or *n*-type regions, respectively. The terminal voltages and currents were labeled as $e_m$ and $j_m$ and were related by the expression $R_H j_m = e_m - e_{m-1}$ $(m = 1, ..., k)$ for CW elements and $R_H j_m = e_m - e_{m+1}$ for CCW elements. The effective resistances and voltages of the circuit between *A* and *B* (Fig. 2 (a)) were modeled for a single polarity of magnetic flux density per simulation due to software constraints. For example, with a positive *B*-field, an *n*-doped EG region was modeled by a CW element, whereas, when *B* was negative, a CCW element was used. The opposite holds true for *p*-type regions. A schematic of the checkerboard device (in one of its measured configurations) is illustrated in Fig. 2 (a) to clarify how the circuit was modeled.

## 3. Results and Discussion

*3.1 Experimental checkerboard devices*

Simulations were performed for numerous configurations of source and drain placement, resulting in the output of various quantized resistances that take the form $R_{AB} = qR_H$, where $R_H$ is the Hall resistance at the $\nu = 2$ plateau ($R_H \approx 12906$ Ω) and *q* is defined as the *coefficient of effective resistance* (CER). The CER is expressed either as an integer or a fraction.

The example schematic of Fig. 2 (a) represents the experimental schematic shown in the inset of Fig. 2 (c). For letters (b) through (e), the quantum Hall devices' total resistances were measured from -9 T to 9 T, with each configuration illustrated in the inset of its corresponding data. Each region contained either two or three available contacts, with some of them being using concurrently for internal voltage measurements. For all data, the temperature was fixed at 1.6 K and a total current of 1 µA was used.

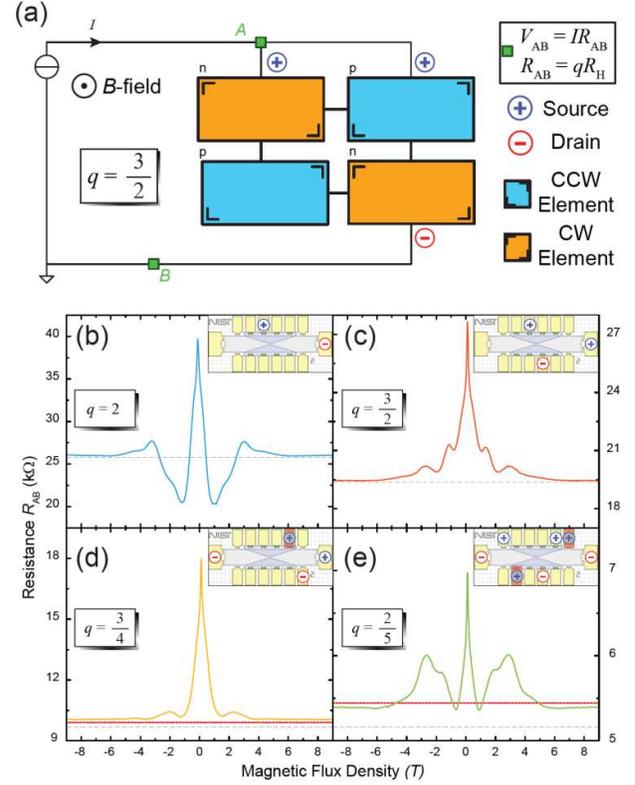

Fig. 2. (a) Schematic of the EG checkerboard *pn*J device from Fig. 1 (a) is shown as part of a circuit intended to output many values of quantized resistance. In this case, two positive current terminals were used (top side) and one negative terminal was used (lower right). (b)-(e) The magnetoresistances of the device between points *A* and *B* are shown as a function of magnetic flux density. LTspice simulation results are marked as gray dashed lines for reference. The red dotted lines in (d) and (e) are results from the LTspice simulation when accounting for contact resistance. Insets show an illustration of each of the measured configurations, with the resulting CERs being highlighted as (b) 2 (c) $\frac{3}{2}$ (d) $\frac{3}{4}$ (e) $\frac{2}{5}$

With sufficiently high magnetic flux densities, typically larger than ± 7 T, the resistance measurements asymptotically approach their predicted values, as represented by dashed gray lines. The CERs of each of the configurations were: 2, $\frac{3}{2}$, $\frac{3}{4}$, and $\frac{2}{5}$. Without any adjustments, these data all agree with their ideal predictions within an error of 2 %, except for (e), which agrees within 5 % error. These errors may be a function of nonideal contact resistance, which would slightly modify the total current allowed to flow into the device from the contact in question and thereby contribute to an error of the measured CER. The observed symmetries with respect to magnetic flux density were also predicted and have been observed in similar instances [25-28].

By having a closer look at the data, the errors may be improved by considering additional measurements and conditions, notably those of possible contact resistance. To begin, Fig. 2 (b) displays a data curve that has been adjusted to account for the following condition: upon starting the experiment, the assumed current of 1 μA was set by using a voltage source of 1 V and a 1 MΩ resistor across a relatively low-resistance Hall configuration at 0 T. The original error arose at high fields, where the measured voltage, including contact resistance of this configuration, was approximately 25.341 mV. Under the original assumption, one could divide this quantity by the known current, which was approximately 1 μA. However, the addition of this quantized resistance to the circuit actually modified the circuit current to about 0.975 μA, which yielded the adjusted displayed resistance plateau valued at about 25.99 kΩ (blue curve). Such consideration yields a final error of 0.7%.

Subsequent measurements were performed with the voltage source adjusted at ± 9 T to provide 1 μA to the circuit more accurately. Resistance measurements were collected at ± 9 T for each of the ten available contacts (transistor outline 8 device package) by using a typical three-terminal method. Since the utilized contacts for Fig. 2 (b) and (c) exhibited low resistance (< 5 Ω), no further adjustments were necessary. The final error for Fig. 2 (c) was therefore 0.5%.

For the case of Fig. 2 (d), one of the measured contacts had a resistance of 1.75 kΩ, which is not negligible. This contact is shown in orange along with a terminal polarity having a gray background rather than white. As mentioned earlier, contacts not used for these data were concurrently used for internal voltage measurements and were thus unavailable for use. Though the ideal CER is represented as a dashed gray line, a second LTspice simulation was necessary to account for the contact resistance since its presence would modify the overall measurable CER. A dotted red line represents the new prediction, which places this configuration at a final error of 1.7%.

In the final case of Fig. 2 (e), the same analysis applied. A second contact had a measured resistance of 2.4 kΩ. By inputting this information into the LTspice simulation, a new value for this circuit was predicted and also represented as a dotted red line. By accounting for these resistances, the error between what was measured and predicted for this configuration was just under 1.0%. The necessity of altering the simulation for imperfect contacts warrants the exploration of using superior alloys for these types of devices, as has been demonstrated in other recent work by using superconducting materials [28, 50].

### 3.2 Proposing the M × M checkerboard

From the experimental point of view, the checkerboard devices whose data constitute Fig. 2 have now been demonstrated to function without the need for performing complex fabrication processes like electron beam lithography or related methods. Rather, a simple ultraviolet lithography process was sufficient to make narrow *pn*Js resulting in propagated edge-states, as seen by the quantized behavior at high magnetic flux density.

With that demonstration, we now turn to propose future devices for outputting quantized resistances in a way that is relatively tunable. The proposed device geometries can be topologically varied for other applications as well (for instance, the Dirac fermion microscope [31] or mesoscopic valley filter [32]), thereby granting conceptual merit to methods that are scalable and devices that are gateless.

For the purpose of obtaining a variety of quantized resistances for a subfield like metrology, suppose that a single source and drain are affixed to the opposite corners of a 2 × 2 device like the one illustrated in Fig. 3 (a), which is pointed to in red from its corresponding predicted CER $q_1$. Here, the subscript denotes the total number of terminals used, $N$, minus one to remain consistent with the literature [25-28]. As the number of regions per side *M* increases to infinity, the CER exhibits convergent behavior. Fig. 3 (a) is one example of how a square checkerboard can be used to select the desired CER within a certain neighborhood of discrete, quantized values.

One device of suitable next-stage development would be the 3 × 3 device shown in Fig. 3 (b). With this geometry of *pn*Js, the parameter space for placement and number of sources and drains becomes much larger and may thus yield a CER more suitable for a particular application. Take note of the CERs listed in a particular color, either black, green, gold, or purple. By modifying the four terminals' positions (or three in the case of the green CER), the obtainable CER varies significantly and may be customized to fit the needs of the user's circuit. In this demonstration, the black, green, gold and purple configurations were found to have CERs of $\frac{10}{13}$, $\frac{134}{99}$, $\frac{41}{30}$, and $\frac{29}{37}$, respectively.

### 3.3 Proposing different region geometries

By utilizing different device geometries via customizable junction placement, one may label such placement as a degree of freedom when it comes to fabricating devices with outputs of a specific quantized resistance. In Fig. 3 (c), a "Zeno's paradox" geometry was simulated whereby the number of single quartile (lower right quartile) subdivisions *M* was increased and each of its corresponding CERs were plotted below in black. This trend appears to linearize as $M$ approaches infinity but initially has some offset from that linear behavior. The difference between the asymptotic behavior and the actual value of the CER were subtracted for each $M$, and the result is plotted in purple on the right vertical axis. From these plots, it is evident that the behavior of the CER with increasing quartile subdivisions rapidly converges to a linear trend with a slope of exactly 3. The physical interpretation of this number is that there are two regions of blue separated by a third region of orange. Such slopes may be engineered to output with different values depending on the careful placement of $pn$Js throughout the device, which need not be symmetric.

example, a "Zeno's paradox" geometry is used whereby the number of quartile subdivisions $M$ is increased, with the corresponding CERs plotted below in black. In purple, the difference between the simulated CER with the asymptotic behavior as $M$ approaches infinity is plotted.

## 4. Conclusion

This work reports the fabrication and testing of millimetre-scale EG checkerboard $pn$J devices and corresponding measurements of such devices in the quantum Hall regime. Experimental data were compared with results from LTspice current simulations and agreed well. Overall, these experiments have both validated the use of these scalable $pn$J devices for quantum electrical circuits as well as substantiated the methods utilized for fabricating large-scale $pn$Js that could potentially be applied to numerous research efforts.

## Acknowledgements and Notes

The work of DKP at NIST was made possible by C-T Liang of National Taiwan University. The work of MM at NIST was made possible by M Ortolano of Politecnico di Torino and L Callegaro of Istituto Nazionale di Ricerca Metrologica, and the authors thank them for this endeavor. The authors would like to express thanks to S Payagala and A Levy for their assistance in the NIST internal review process.

Commercial equipment, instruments, and materials are identified in this paper in order to specify the experimental procedure adequately. Such identification is not intended to imply recommendation or endorsement by the National Institute of Standards and Technology or the United States government, nor is it intended to imply that the materials or equipment identified are necessarily the best available for the purpose.

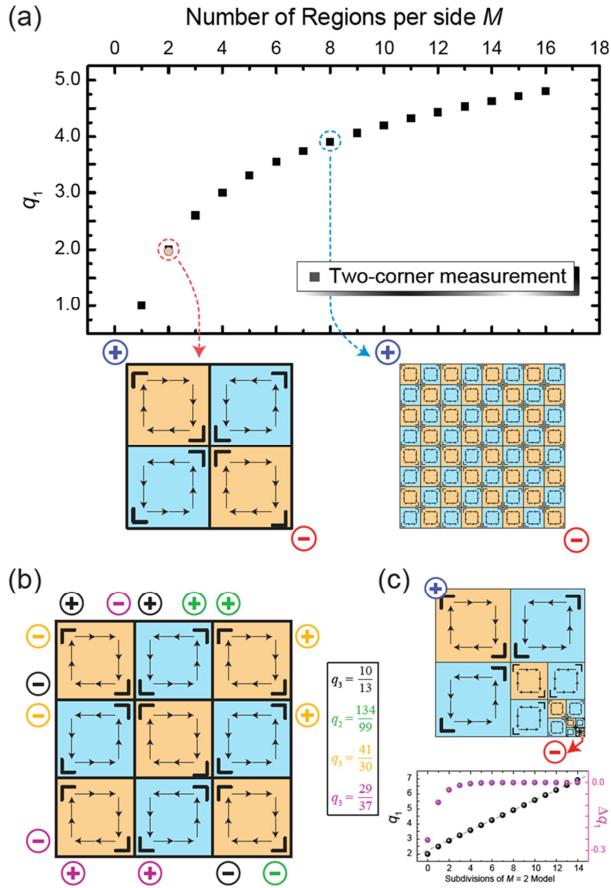

Fig. 3. (a) Simulated data representing the CERs of the two-terminal measurements for checkerboard $pn$J devices of varying number of distinct regions per side, $M$. The source and drain were placed at opposite ends of the checkerboard device for each simulation. The accompanying illustrations show two example cases, the 2 × 2 checkerboard (done experimentally) and the 8 × 8 checkerboard, with a corresponding dashed blue line eminating from its data point. (b) The 3 × 3 checkerboard configuration is explored in more detail to highlight the variability in accessible quantized resistances. CERs are color coded to match their corresponding source and drain configurations. (c) A device is proposed here as a means to exemplify the impact of using various geometries as a degree of freedom when constructing devices with outputs of specific quantized resistances. In this